# Percolation of optical excitation mediated by near-field interactions

Makoto Naruse[1], Song-Ju Kim[2], Taiki Takahashi[3], Masashi Aono[4,5], Kouichi Akahane[1],
Mario D'Acunto[6], Hirokazu Hori[7], Lars Thylén[8,9], Makoto Katori[10] and Motoichi Ohtsu[11]

[1] *Network System Research Institute, National Institute of Information and Communications Technology, 4-2-1 Nukui-kita, Koganei, Tokyo 184-8795, Japan*

[2] *WPI Center for Materials Nanoarchitectonics, National Institute for Materials Science, 1-1 Namiki, Tsukuba, Ibaraki 305-0044, Japan*

[3] *Department of Behavioral Science, Center for Brain Science, Center for Experimental Research in Social Sciences, Hokkaido University, Sapporo 060-0808, Japan*

4 *Earth-Life Science Institute, Tokyo Institute of Technology, 2-12-1 Ookayama, Meguru-ku, Tokyo 152-8550, Japan*

5 *PRESTO, Japan Science and Technology Agency, 4-1-8 Honcho, Kawaguchi-shi, Saitama 332-0012, Japan*

[6] *Institute of Structure of the Matter, Italian National Research Council (CNR), Via Moruzzi 1, 56124, Pisa, Italy*

[7] *Interdisciplinary Graduate School of Medicine and Engineering, University of Yamanashi, Takeda, Kofu, Yamanashi 400-8511, Japan*

[8] *Department of Theoretical Chemistry, Royal Institute of Technology (KTH), S-106 91 Stockholm, Sweden*

[9]*Hewlett-Packard Laboratories, 1501 Page Mill Rd., Palo Alto, CA 94304 USA*

[10] *Department of Physics, Faculty of Science and Engineering, Chuo University, 1-13-27 Kasuga, Bunkyo-ku, Tokyo 112-8551, Japan*

[11] *Department of Electrical Engineering and Information Systems, Graduate School of Engineering, The University of Tokyo, 2−11−16 Yayoi, Bunkyo−ku, Tokyo 113−8656, Japan*

**Abstract:** Optical excitation transfer in nanostructured matter has been intensively studied in various material systems for versatile applications. Herein, we discuss the percolation of optical excitations in randomly organized nanostructures caused by optical near-field interactions governed by Yukawa potential in a two-dimensional stochastic model. The model results demonstrate the appearance of two phases of percolation of optical excitation as a function of the localization degree of near-field interaction. Moreover, it indicates sublinear scaling with percolation distance when the light localization is strong. The results provide fundamental insights into optical excitation transfer and will facilitate the design and analysis of nanoscale signal-transfer characteristics.

Optical excitation transfer has been intensively studied various material systems [1-3] and utilized in versatile applications including nanobiosensors [4], solid-state lighting [5], signal conversion [6], optical switching [7], and intelligent functions [8]. The theory of optical excitation transfer has been explained by local optical near-field interactions, which describe optical excitation transfer involving conventionally dipole-forbidden transitions [9, 10].

In experimental efforts, one critical concern is to regulate the sizes and positions of nanostructures so that optical near-field interactions are induced between them to obtain the desired functions. Thus, it is necessary to model nanophotonic devices and systems composed of multiple nanostructures arranged in varying configurations to characterize and design designated functions. In a previous study, we constructed a stochastic model to examine optical excitation transfer in multilayer quantum dot (QD) devices whereby the variation in QD size and temperature-dependent energy band broadening are concerned in a unified manner [11]. However, the spatial inhomogeneity was not considered and a better fundamental understanding needs to be developed; basic phenomena such as the percolation of optical excitation in random media have not yet been examined. Nomura *et al.* demonstrated long-range optical excitation transfer in randomly distributed core-shell QDs [12]; such a system has been successfully utilized in intelligent devices such as those for applications including decision making [8]. However, the performance limitations, fundamental characteristics (e.g., robustness), and systematic design methodologies of these systems have not yet been clarified; hence, further insights into optical excitation transfer are required.

In this paper, we characterize the percolation behaviour of optical excitation related to near-field interactions governed by Yukawa-type potential in a randomly organized nanoparticle system distributed on a two-dimensional system. This perspective of percolation provides interesting insights in a broad range of scientific disciplines such as physics, materials science, and complex networks [13,14]. Herein, percolation refers to the optical excitation transfer from a source node to a sink node. By intentionally destructing internal material systems between these nodes, (i.e., deleting some elemental structures from the original system), we examine how the optical excitation transfer from the source to sink node is altered by taking into the effects of optical near-field interactions. We demonstrate that two different types of percolation appear depending on the degree of localization of the optical near fields. Furthermore, we show that the distant-dependent percolation deviates from normal linear scaling when the light localization is strong.

We begin by reviewing some of the basic theoretical elements of optical excitation transfer mediated by near-field interactions [9,15]. We assume two spherical QDs with radii $R_S$ and $R_L$ (termed as $QD_S$ and $QD_L$, respectively) located in close proximity (Figure 1(a)). The energy eigenvalues of the states specified by quantum numbers $(n,l)$ are given by

$$E_{nl} = E_g + E_{ex} + \frac{\hbar^2 \alpha_{nl}^2}{2MR^2} \quad (n=1,2,3,\cdots), \tag{1}$$

where $E_g$ is the band gap energy of the bulk semiconductor, $E_{ex}$ is the exciton binding energy in the bulk system, and $M$ is the effective mass of the exciton. $\alpha_{nl}$ are determined from the boundary conditions such as $\alpha_{n0} = n\pi$ and $\alpha_{11} = 4.49$. According to equation (1), the energy level of quantum number (1,0) in $QD_S$ and that of quantum number (1,1) in $QD_L$ are resonant with each other if $R_L / R_S = 4.49/\pi \approx 1.43$. Note that the optical excitation of the (1,1)-level in $QD_L$ corresponds to an electric dipole-forbidden transition. An optical near field, denoted by $U$ in figure 1(a), given by the Yukawa-type potential

$$U^{-1} = \frac{\exp(-\mu r)}{r} \tag{2}$$

allows this level to be populated due to the steep electric field in the vicinity of $QD_S$ [9]. Here, $r$ is the interdot distance and $\mu$ quantifies the degree of light localization. Therefore, an exciton in the (1,0)-level in $QD_S$ could be transferred to the (1,1)-level in $QD_L$. In $QD_L$, the excitation undergoes energy dissipation by intersublevel relaxation denoted by $\Gamma$, which is faster than the rate of the interdot optical near-field interaction, and the excitation causes a transition to the (1,0)-level and radiation into the far field. Finally, we observe unidirectional optical excitation transfer from $QD_S$ to $QD_L$. Here, we call $QD_S$ the *source* node, whereas $QD_L$ is referred to as the *sink* node.

The model shown in figure 1(a) can be extended to a system composed of *multiple* $QD_S$s and a single $QD_L$, wherein optical excitation generated at the source is transferred to the sink via multiple intermediate $QD_S$s; such systems have been experimentally demonstrated in randomly distributed CdSe/ZnS core-shell QDs [8,12] and InAlAs multilayer QDs formed in Stranski–Krastanov mode [3,11].

We introduce a stochastic model in which QDs are randomly distributed in a rectangular-shaped area; this model is schematically shown in figure 1(b). The radii of $QD_S$ and $QD_L$ are 5 and 7 nm, respectively, and the source $QD_S$ and the sink $QD_L$ are separated by 400 nm. The rectangular-shaped area is 500 nm in the horizontal direction (X-axis) and 50 nm in the vertical direction (Y-axis). Letting the left, lower corner be the Cartesian origin, the source QD is located at $(50, 25)$, and the sink QD is located at $(450, 25)$. The centre positions of the intermediate $QD_S$s are determined by random numbers so that they fall into the rectangular area but outside the areas occupied by other QDs.

We use the following strategy to quantify the signal transfer from the source to sink. First, we identify all of the QDs in the system with the index *i* taking integer values ranging from 1 to *N*, with *N* being the total number of QDs in the system. The distance between QD *i* and QD *j* is denoted by $d_{ij}$. We then introduce the *effective distance* between QD *i* and QD *j* concerning near-field interaction between them defined by the inverse of equation (2), namely, $(\exp(-\mu d_{ij})/d_{ij})^{-1}$. This leads to an $N \times N$ matrix in which element $ij$ represents the effective distance between QD *i* and *j*. We derive the path *K* from the source to the sink such that the total sum of effective distances along the path is minimized. This sum is given by

$$\sum_{(i,j)\in K}\left(\frac{\exp(-\mu d_{ij})}{d_{ij}}\right)^{-1}. \tag{3}$$

Based on equation (3), the shortest path is calculated from the matrix defined above and by using Dijkstra's methods [16].

We intentionally degrade the system by removing some of the intermediate QD$_S$ and then examine how the total effective distance given by equation (3) varies. For example, figures 1(c) and (d) denote instances of systems in which the numbers of removed QDs are 50 and 100, respectively. Meanwhile, the total effective distance depends on $\mu$ (equation (2)), which describes the strength of the light localization. In the numerical evaluation, we prepare 10 different initial QD distributions, each of which experiences 100 different removal patterns for each of number of removed QD. We then evaluate the resultant mean value normalized by the value of equation (3) in the case of zero QD removal, which is hereafter referred to as *effective transmission efficiency (ETE)*.

Figure 2(a) summarizes the results. The localization strength $\mu$ is given by 1, 1/2, 1/3, 1/4, 1/5, 1/6, 1/7, 1/8, 1/9, 1/10, 1/20, 1/30, 1/40, 1/50, 1/60, 1/70, 1/80, 1/90, and 1/100. When localization is strong (e.g., $\mu=1$, indicated by the red curve in figure 2(a)), ETE rapidly degrades to zero by the removal of a low number of QDs. However, when localization is weak (e.g., $\mu=1/100$, denoted by the green curve in figure 2(a)), ETE does not decrease rapidly with increasing QD removal ratio. We observe two different types of ETE curves; one curve is convex downward as a function of QD removal ratio, whereas the other is convex upward (depicted by the red and green ring marks in figure 2(a), respectively). To quantitatively examine the difference, we evaluate the figure-of-merit (FoM), which is defined as the mean value of the second-order derivative of ETE for each of the ETE curves (figure 2(b)). FoM is positive and negative when $\mu$ is greater and lower than approximately 1/20, respectively.

This result can be explained by the appearance of *two phases* of percolation. In one phase, the area $\mu$ is larger than a certain threshold (approximately 1/20), and signal transmission is more *near-field* dominated; thus, the percolation from the source to the sink is easily prohibited by a marginal deconstruction of the internal systems. In the other domain, $\mu$ is lower than the threshold (approximately 1/20), and the signal transmission is more *far-field* dominated; thus, the percolation from the source to the sink is induced, even when the internal structures are heavily degraded.

This property is considered further in the following discussion. Suppose that the interaction function is given by $U^{-1} = 1/r$, which describes the nature of a propagating wave, instead of by equation (2). The effective distance, formerly given by equation (3), is simply reformulated as $\sum d_{ij}$, which yields its minimum value when the source and the sink are directly connected. Hence, different phases of percolation with the Yukawa-type potential (figure 2) never emerge when the potential is given by $U^{-1} = 1/r$. Actually, any interaction function in the form of $1/r^n$ leads to the minimum effective distance from the source to the sink by the *direct* path between the two; therefore, different types of percolation cannot be induced.

As mentioned above, the transition from negative to positive FoM occurs at approximately $\mu = 1/20$. From equation (2), the inverse of $\mu$ takes the unit of size. Because the diameter of the small QD is 10 nm, the transition happens at two times the size of the QD, which is consistent with the general experimental results observed in near-field optical studies demonstrating that the effect of near-field light is comparable to the size of the nanostructure under study [17-19]. Meanwhile, the percolation governed by near-field interactions indicates that the long-range excitation transfer can persist if the internal nanostructure does not contain arranged domains that are sparsely distributed in space, namely, the condition of small *node removal ratio* with respect to the horizontal axis of figure 2(a) is fulfilled. This is consistent with the experimental results of by Nomura *et al*. [12], wherein optical excitation transfer of up to micrometer scale was successfully observed in *densely* organized core-shell QDs.

Now, we investigate the dependence of percolation properties on the distance from source to sink. Supposing that $\mu = 1$, we examine four different positions for the sink QD (figure 3(a)). The source QD is located at *S* = (50, 25), whereas the sink QD is located either at $G_1$ = (150,25), $G_2$ = (250,25), $G_3$ = (350,25), or $G_4$ = (450,25); therefore, the distances between the source *S* and sinks $G_1$–$G_4$ are 100, 200, 300, and 400 nm, respectively.

Figures 3(b)–(e) show examples of the minimum paths *S* to $G_1$, $G_2$, $G_3$, and $G_4$, respectively. For example, the minimum path from *S* to $G_2$, does *not* necessarily overlap with the minimum path from *S* to $G_1$. The same argument is also applicable to the paths {*S* to $G_3$} and {*S* to $G_4$}. Let the effective distance from *S* to $G_1$ be $T_1$. If the effective distance scales *normally* with the physical distance, the effective distance from S to $G_2$ should be $2 \times T_1$ since the physical distance is doubled. In reality (red *x* marks in figure 3(f)), the effective distances from *S* to $G_i$ ($i = 1, \cdots, 4$) are *smaller* than the normally interpolated evaluations (such as $2 \times T_1$ for $G_2$) depicted by the blue circles in figure 3(f). This indicates that the percolation of optical excitation follows a *sublinear* scaling. The relative deviation from the normal scaling is evaluated as a function of the strength of localization $\mu$, as shown in figure 3(g). The deviation approaches unity, indicating that scaling is nearly normal, and under strong light localization, the localization parameter $\mu$ decreases, whereas the deviation increases. This is consistent with the previous results indicating that the light transmission is more far-field dominated at lower $\mu$. As previously discussed, if the interaction is governed by $U^{-1} = 1/r$, which corresponds to far-field light, the effective distance is simply determined by the straight line from the source to the sink; consequently, the scaling is normal.

In summary, we investigated the percolation of optical excitation in randomly organized nanostructures mediated by near-field light characterized by the Yukawa-type potential. The model results clearly demonstrate two different phases of percolation; the percolation is easily blocked by a slight internal structural degradation under strong light localization, whereas weaker light localization provides robust percolation from the source QD to the sink QD. Furthermore, the excitation transfer efficiency exhibits sublinear scaling with respect to the actual physical distance from the source to sink node, especially in the case when light localization is strong. This study contributes fundamental insights into the design and analysis of nanophotonic devices and systems based on multiple nanostructures.

**Acknowledgments**

This work was supported in part by the Grant-in-Aid for Scientific Research and the Core-to-Core Program, A. Advanced Research Networks from the Japan Society for the Promotion of Science.

**Figure captions**

**Figure 1**. (a) Optical excitation transfer from the smaller dot (source dot) to the larger one (sink dot) via optical near-field interactions. (b) Model of randomly distributed quantum dots distributed in a two-dimensional structure. We evaluate the transmission efficiency on the basis of the path that minimizes the effective distance governed by the near-field potential from the source to sink dot (the red solid line depicts the minimum path.) (c,d) Paths that minimize the effective distance when some of the internal nanostructures are removed from the system; 50 (c) and 100 (d) particles are removed from the original systems in (b) and (c), respectively.

**Figure 2.** (a) Effective transmission efficiency (ETE) from the source to sink as a function of node-removal ratio. The ETE curves are convex downward and upward with strong and weak light localization, respectively. (b) The average value of the second-order derivative of ETE curves are evaluated as a function of light localization strength.

**Figure 3.** (a) Distance-dependent percolation. (b-e) Paths that minimize the effective distance from the source to (b) sink 1, (c) sink 2 (d) sink 3, and (e) sink 4. The paths do not necessarily overlap each other. (f, g) Evaluation of distance-dependent percolation of optical excitation with normal scaling. When the light localization is strong, the distance dependencies exhibit sublinear scaling.

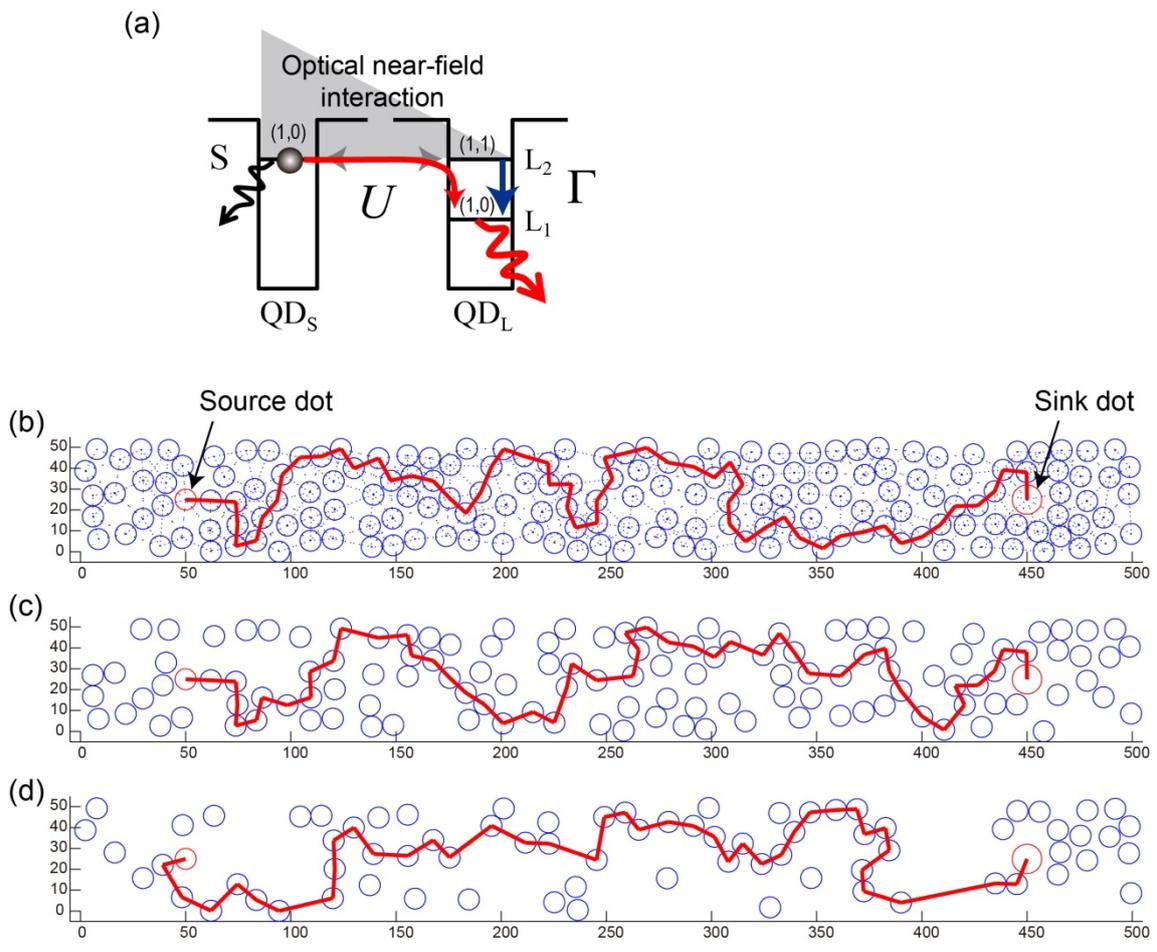

Figure 1

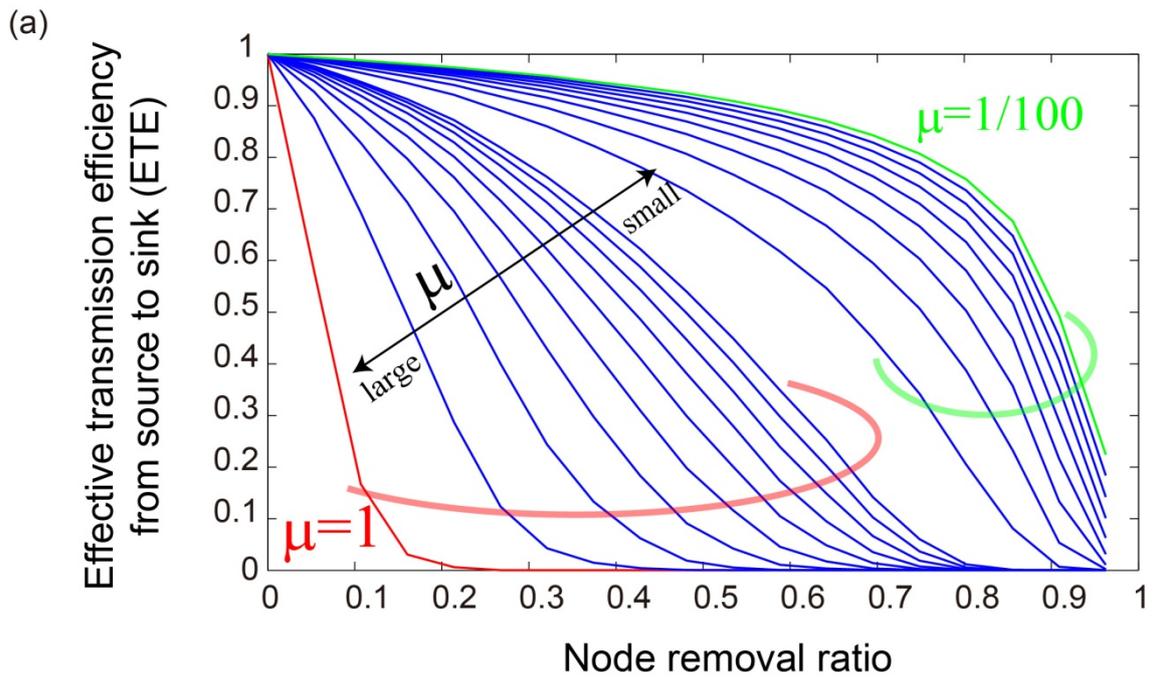

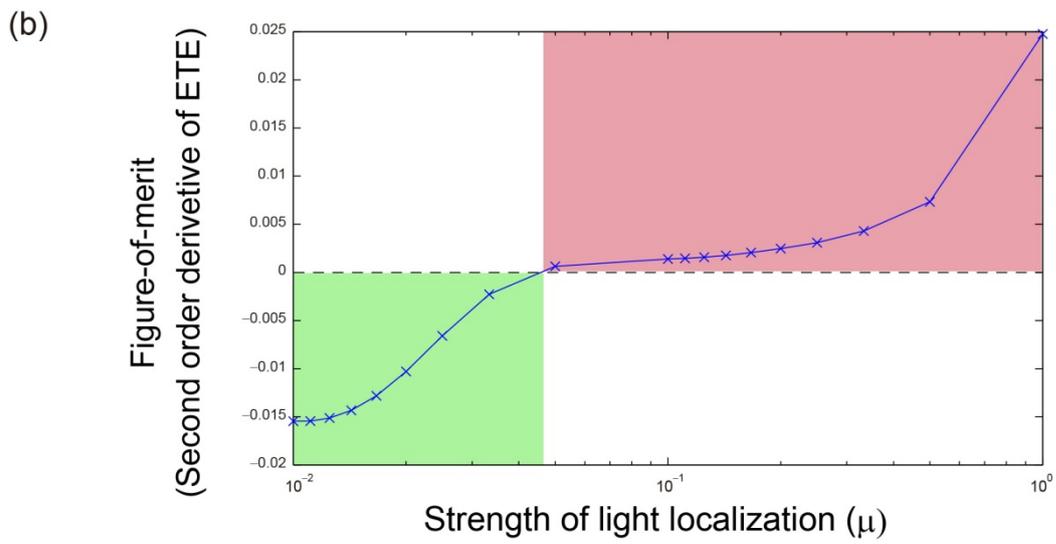

Figure 2

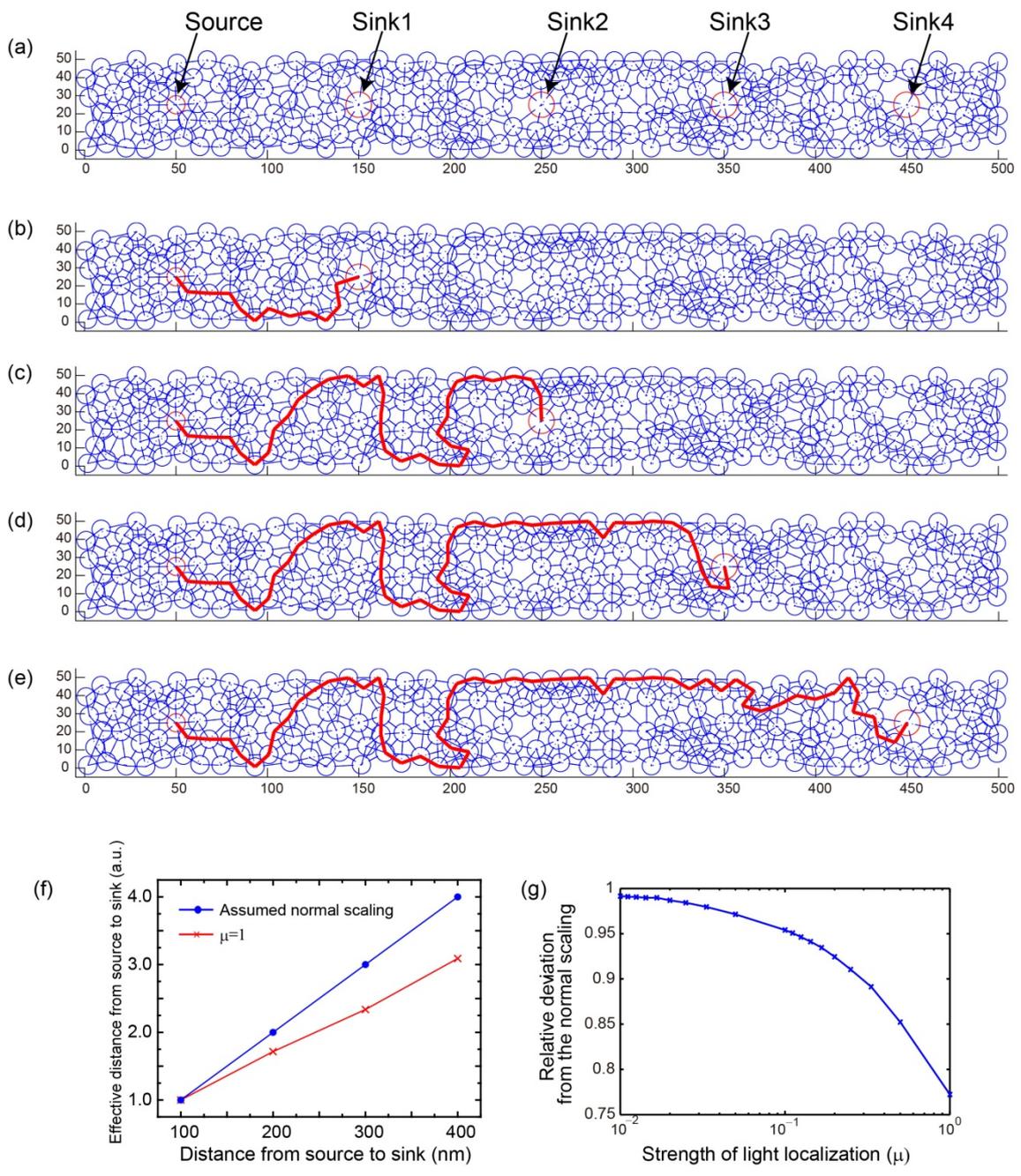

Figure 3